\documentclass[letter]{aa} 

\usepackage{graphicx,txfonts,natbib}
\bibpunct{(}{)}{;}{a}{}{,} 
\usepackage{color}
\begin{document}

   \title{Spectroscopic follow-up of L- and T-type proper- motion member candidates in the Pleiades}


   \author{M$.$ R$.$ Zapatero Osorio\inst{1}
          \and
                V$.$ J$.$ S$.$ B\'ejar\inst{2,3}
          \and
                E$.$ L$.$ Mart\'in\inst{1}
          \and
                M$.$ C$.$ G\'alvez Ortiz\inst{1}
          \and
                R$.$ Rebolo\inst{2,3}
          \and
                G$.$ Bihain\inst{4}
          \and
                Th$.$ Henning\inst{5}
          \and
                S$.$ Boudreault\inst{6}
          \and
                B$.$ Goldman\inst{5}
          \and
                R$.$ Mundt\inst{5}
          \and
                J$.$ A$.$ Caballero\inst{7}
          \and
                P. A. Miles-P\'aez\inst{2,3}
          }

   \institute{Centro de Astrobiolog\'\i a (CSIC-INTA), Carretera de Ajalvir km 4, E-28850 Torrej\'on de Ardoz, Madrid, Spain\\
              \email{mosorio@cab.inta-csic.es}
         \and
               Instituto de Astrof\'\i sica de Canarias, V\'\i a L\'actea s/n, E-38205 La Laguna, Tenerife, Spain
         \and
               Dept$.$ Astrof\'\i sica, Universidad de La Laguna, E-38206 La Laguna, Tenerife, Spain
         \and
               Leibniz-Institut f\"ur Astrophysik Potsdam (AIP), An der Sternwarte 16, 14482, Potsdam, Germany
         \and
               Max-Planck-Institut f\"ur Astronomie, K\"onigstuhl 17, 69117 Heidelberg, Germany
         \and
               GEPI, Observatoire de Paris, CNRS, Universit\'e Paris Diderot; 5 Place Jules Janssen, F-92190 Meudon, France
         \and
               Centro de Astrobiolog\'\i a (CSIC-INTA), PO Box 78, E-28691 Villanueva de la Ca\~nada, Madrid, Spain
             }

   \date{Received ; accepted }
 
  \abstract{We report on the near-infrared ($JHK$-bands) low-resolution spectroscopy and red optical ($Z$-band) photometry of seven proper-motion, very low-mass  substellar member candidates of the Pleiades cluster with magnitudes in the interval $J$\,=\,17.5--20.8 and $K$\,=\,16.1--18.5 mag. Spectra were acquired for six objects with the LIRIS and NIRSPEC instruments mounted on the 4.2-m William Herschel and the 10-m Keck II telescopes, respectively. $Z$-band images of two of the faintest candidates were collected with the ACAM instrument on the WHT. The new data confirm the low temperatures of all seven Pleiades proper motion candidates. From the imaging observations, we find extremely red $Z-J$ and $Z-K$ colors that suggest that the faintest target, Calar Pleiades~25, has a Galactic rather than extragalactic nature. We tentatively classify the spectroscopic targets from early-L to $\sim$T0 and suggest that the L/T transition, which accounts for the onset of methane absorption at 2.1 $\mu$m, may take place at $J \approx 20.3$ and $K \approx 17.8$ mag in the Pleiades (absolute values of $M_J \approx 14.7$ and $M_K \approx 12.2$ mag). We find evidence of likely low-gravity atmospheres based on the presence of triangular-shape $H$-band fluxes and the high flux ratio $K/H$ (compatible with red $H-K$ colors) of Calar Pleiades~20, 21, and~22, which is a feature also seen in field low-gravity dwarfs. Weak K\,{\sc i} absorption lines at around 1.25 $\mu$m are probably seen in two targets. These observations add support to the cluster membership of all seven objects in the Pleiades. The trend delineated by the spectroscopic sequence of Pleiades late-M and L dwarfs resembles that of the field. With masses estimated at 0.012--0.015 M$_\odot$ (solar metallicity and 120 Myr), Calar Pleiades~20 (L6$\pm$1), 21 (L7$\pm$1), and~22 (L/T) may become the coolest and least massive Pleiades members that are corroborated with photometry, astrometry, and spectroscopy. Calar Pleiades~25 ($<$0.012 M$_\odot$) is a firm free-floating planetary-mass candidate in the Pleiades. }


   \authorrunning{Zapatero Osorio et al$.$}
   \titlerunning{Spectroscopic follow-up of Pleiades proper motion candidates}

   \maketitle

%

\small

\section{Introduction}
The Pleiades star cluster offers a unique opportunity to scrutinize brown dwarfs and planetary-mass objects of known (intermediate) age and metallicity. With a solar abundance \citep{soderblom09}, an age of 120\,$\pm$\,10 Myr \citep{basri96,martin98,stauffer98}, and a distance of 133.5\,$\pm$\,1.2 pc \citep{soderblom05,melis14}, the Pleiades is a rich cluster containing over 1000 members \citep{sarro14}. Previous attempts to identify Pleiades substellar objects with L and T types were reported by \citet{bihain06}, \citet{casewell07,casewell10}, and \citet{lodieu12}. Spectroscopic follow-up observations have confirmed L-type Pleiades brown dwarf members down to $\sim$L4 spectral type \citep{bihain10}, yet no T-type cluster object has been unambiguously identified to date (see \citealt{casewell11,lucas13}), partly because these sources are intrinsically very dim and spectroscopic observations turn out to be challenging. At the age of the Pleiades, L and T dwarfs are expected to have masses in the interval 0.005--0.040 M$_\odot$ \citep{chabrier00}. At much younger ages ($\le$10 Myr), objects of these masses, which are typically classified as late-Ms and Ls, populate star-forming regions such as Orion and Upper Scorpius (see review by \citealt{luhman12rev}). Their number suggests a rising mass spectrum, meaning that the count of substellar objects per mass unit softly increases with low mass (see \citealt{bayo11,pena12}). To construct the picture of the complete evolutionary path described by low-mass brown dwarfs and planetary-mass objects from warm temperatures until they become ultra-cool Y-type dwarfs, the Pleiades must be
investigated deeply. These studies will also lead to the derivation of a reliable cluster mass function, which is a key ingredient for the models of stellar and substellar formation, including planetary systems. Furthermore, Pleiades low-mass brown dwarfs and planetary-mass objects are also interesting because they may  provide additional insight into  the physical properties and evolution of giant planets that orbit stars. 

In \citet{osorio14a} we discovered a population of 19 faint red candidates in an area of 0.8 deg$^2$ whose proper motions are compatible with the distinctive one of the Pleiades. According to their optical, near-, and mid-infrared colors, they could be L and early-T dwarfs. If corroborated as cluster members, their masses would span the range 0.008--0.072 M$_\odot$. Here, we report on the photometric and spectroscopic follow-up of seven of these Pleiades proper motion candidates with $J$ and $K$ magnitudes in the intervals 17.5--20.8 and 16.1--18.5 mag: Calar Pleiades~16, 18--22, and~25. (From now on, we use abridged names for the targets.) Our main objective is to address the candidates' membership in the Pleiades, which would allow us to characterize the spectroscopic properties of the Pleiades brown dwarfs and the free-floating planetary mass objects and to determine a robust substellar mass function.

\section{Observations \label{obs}}
\begin{figure*}
   \centering
   \includegraphics[width=0.37\textwidth]{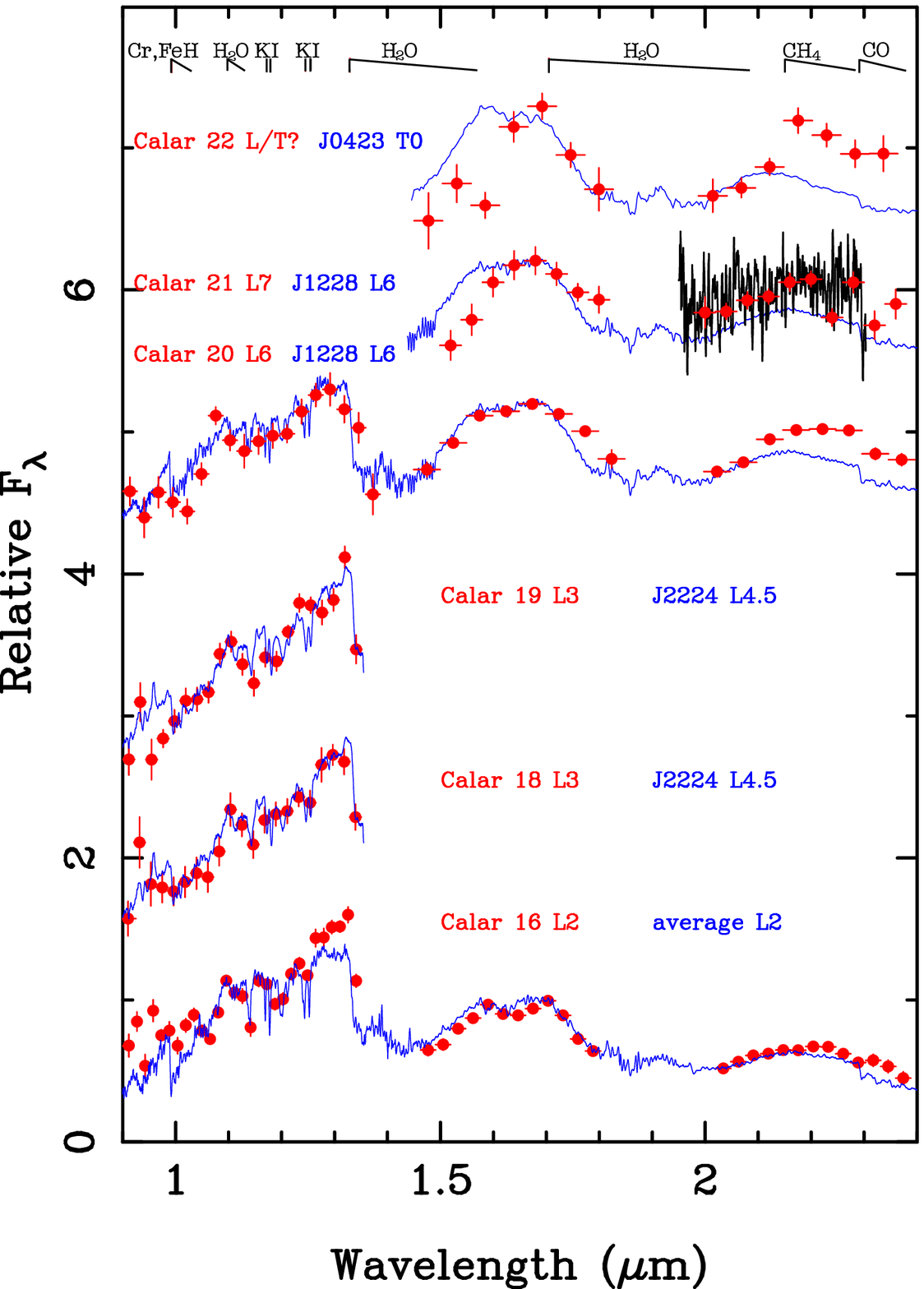}
   \includegraphics[width=0.37\textwidth]{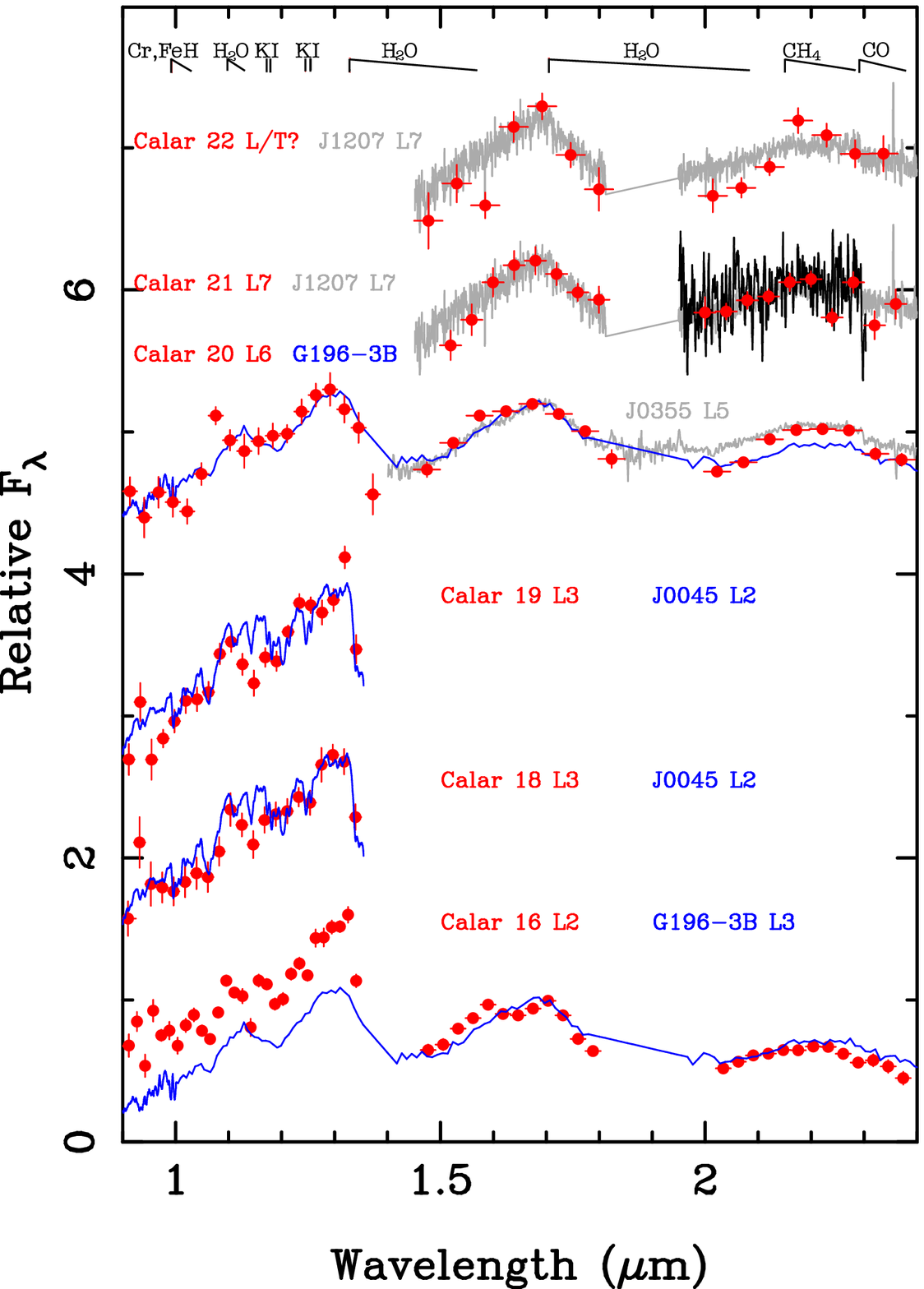}
   \caption{LIRIS (red dots) and NIRSPEC (black line) spectra of Pleiades proper motion candidates. The NIRSPEC spectrum of Calar~21 is normalized to the $K$-band LIRIS spectrum. The {\sl left panel} illustrates the comparison of the Pleiades data with field, high-gravity dwarfs, and the {\sl right panel} highlights the comparison with young (low-gravity) sources. The reference spectra (blue and gray lines) correspond to 2MASS\,J22244381$-$0158521 (L4.5, LIRIS data), DENIS\,J122815.2$-$154733 (L6), and SDSS\,J042348.57$-$041403.5 (T0) in the left panel, and to G\,196-3B (L3), 2MASS\,J00452143$+$1634446 (L2, LIRIS data), 2MASS\,J03552327$+$1133437 (L5, LIRIS data), and 2MASS\,J12073347$-$3932540b (L7) in the right panel. To improve the quality of the Pleiades data, the following binning factors (in pixels) were applied: 25 ($ZJ$) and 29 ($HK$) for Calar~16, 35 ($ZJ$) for Calar~18 and~19, 44 ($ZJ$) and 51 ($HK$) for Calar~20, 41 ($HK$) for Calar~21, and 55 ($HK$) for Calar~22. The vertical error bars of the red dots stand for the errors of the mean fluxes, and the horizontal bars account for the wavelength coverages of the binning intervals. For clarity, regions of strong telluric absorption were intentionally removed from the LIRIS and NIRSPEC spectra, and the data normalized to 1.0 at around 1.3 or 1.68 $\mu$m and shifted by a constant. }
              \label{spectra}
    \end{figure*}

We collected low-resolution near-infrared spectra of six faint Pleiades proper motion candidates (Calar~16 and 18--22, \citealt{osorio14a}) using the Long-slit Intermediate Resolution Infrared Spectrograph, LIRIS \citep{manchado04}, at the Cassegrain focus on the 4.2-m William Herschel Telescope (WHT) and the echelle and grating Near-Infrared Spectrometer, NIRSPEC \citep{mclean98}, at the Nasmyth focus on the 10-m Keck\,II telescope. Sloan $z$-band images of Calar~21 and~25 were acquired with the Auxiliary-port Camera (ACAM) at the folded-Cassegrain focus on the WHT. Together with the science targets, we also observed known spectroscopic and photometric L-type reference dwarfs. Table~\ref{obslog} provides the journal of the spectroscopic and imaging observations including abridged objects names, observing dates in Universal Time (UT), integration times, and air masses. In Sect$.$~\ref{appendix} (online material), we provided detailed description of the observing strategy, instrumental configuration, data reduction, and description of the reference L-type sources. The LIRIS $ZJ$ (0.9--1.35 $\mu$m) and $HK$ (1.45--2.4 $\mu$m) spectra and the NIRSPEC $K$-band (1.95--2.31 $\mu$m) spectra have resolutions of 25~\AA, 39 \AA, and 17 \AA, respectively. The Sloan photometry was converted into the UKIDSS system using the AB to Vega correction and the color terms appropriate for mid- to late-L types given by \citet{hewett06}. Calar~25 remains undetected after 1.4-h on source integration time.  In Table~\ref{obslog}, we list a 4-$\sigma$ $Z$-band upper limit of 23.35 mag ($\sim$3 mag deeper than the UKIDSS survey, \citealt{lodieu12}), where $\sigma$ stands for the sky noise at the position expected for the source. Figure~\ref{spectra} illustrates the resulting LIRIS and NIRSPEC spectra of Pleiades candidates and reference L dwarfs. Given the faint magnitudes of our science targets, the spectra have a poor signal-to-noise ratio (S/N = 2.5--8). To improve the data quality, we depict binned spectra as indicated in the figure caption. For a proper scaling of the $ZJ$ and $HK$ separate spectra of Calar~16 and~20, we employed the objects' $J$ and $H$ photometry of \citet{osorio14a}.

\begin{table}
\caption{Observing log. \label{obslog}}
\centering
\scriptsize
\tabcolsep=0.1cm
\begin{tabular}{lclrccc}
\hline\hline
Object & $J$\tablefootmark{a}   & Instrument\tablefootmark{b} & Exposure & UT & Airmass & SpT/$Z$\tablefootmark{c} \\
       & (mag) &                             & (s)      &    &         &  (mag) \\
\hline
Calar~16                      & 17.54 & LIRIS/$ZJ$ & 14$\times$600 & 2013 Oct 12 & 1.01 & L2\,$\pm$\,1 \\
                              &       & LIRIS/$HK$ &  4$\times$500 & 2014 Feb 21 & 1.07 & L2\,$\pm$\,1 \\
Calar~18                      & 18.18 & LIRIS/$ZJ$ & 12$\times$600 & 2013 Oct 12 & 1.39 & L3\,$\pm$\,1 \\
Calar~19                      & 18.49 & LIRIS/$ZJ$ & 10$\times$900 & 2013 Oct 13 & 1.04 & L3\,$\pm$\,1 \\
Calar~20                      & 18.63 & LIRIS/$ZJ$ & 10$\times$600 & 2013 Oct 11 & 1.02 & L6\,$\pm$\,1 \\
                              &       & LIRIS/$HK$ & 12$\times$500 & 2014 Feb 20 & 1.06 & L6\,$\pm$\,1 \\
Calar~21                      & 20.23 & NIRSPEC/$K$&  4$\times$600 & 2013 Dec 13 & 1.02 & L7\,$\pm$\,1 \\
                              &       & LIRIS/$HK$ & 16$\times$500 & 2014 Feb 18 & 1.10 & L7\,$\pm$\,1 \\
                              &       & ACAM/$z$   & 17$\times$180 & 2014 Jan 15 & 1.05 & 22.91\,$\pm$\,0.15 \\
Calar~22                      & 20.29 & LIRIS/$HK$ & 18$\times$500 & 2014 Feb 19 & 1.15 & L/T \\
Calar~25                      & 20.83 & ACAM/$z$   & 42$\times$120 & 2014 Feb 21 & 1.26 & $\ge$23.35\tablefootmark{d} \\
\hline
J0045$+$1634\tablefootmark{e} & 13.06 & LIRIS/$ZJ$ &  4$\times$300 & 2013 Oct 10 & 1.19 & L2\tablefootmark{f} \\
J0355$+$1133\tablefootmark{e} & 14.05 & LIRIS/$HK$ &  2$\times$300 & 2014 Feb 18 & 1.07 & L5\tablefootmark{f} \\
J0908$+$5032                  & 14.55 & ACAM/$z$   &   3$\times$30 & 2014 Feb 21 & 1.09 & L5/L9\tablefootmark{f} \\
J2224$-$0158                  & 14.07 & LIRIS/$ZJ$ &  2$\times$400 & 2013 Oct 12 & 1.42 & L4.5\tablefootmark{f} \\
\hline
\end{tabular}
\tablefoot{ 
\tablefoottext{a}{UKIDSS photometric system for the Calar objects, 2MASS for the field dwarfs below the line.}
\tablefoottext{b}{Filter (ACAM imaging) or grism (LIRIS spectroscopy).}
\tablefoottext{c}{UKIDSS photometric system (Vega magnitudes).}
\tablefoottext{d}{4-$\sigma$ detection limit.}
\tablefoottext{e}{Young field dwarf \citep{cruz09,osorio14b}.}
\tablefoottext{f}{Spectral type  from \citet{kirk00} and \citet{cruz03,cruz09}.}
}
\end{table}

\section{Results and discussion}
\begin{figure}
   \centering
   \includegraphics[width=0.24\textwidth]{zjk.ps}
   \includegraphics[width=0.24\textwidth]{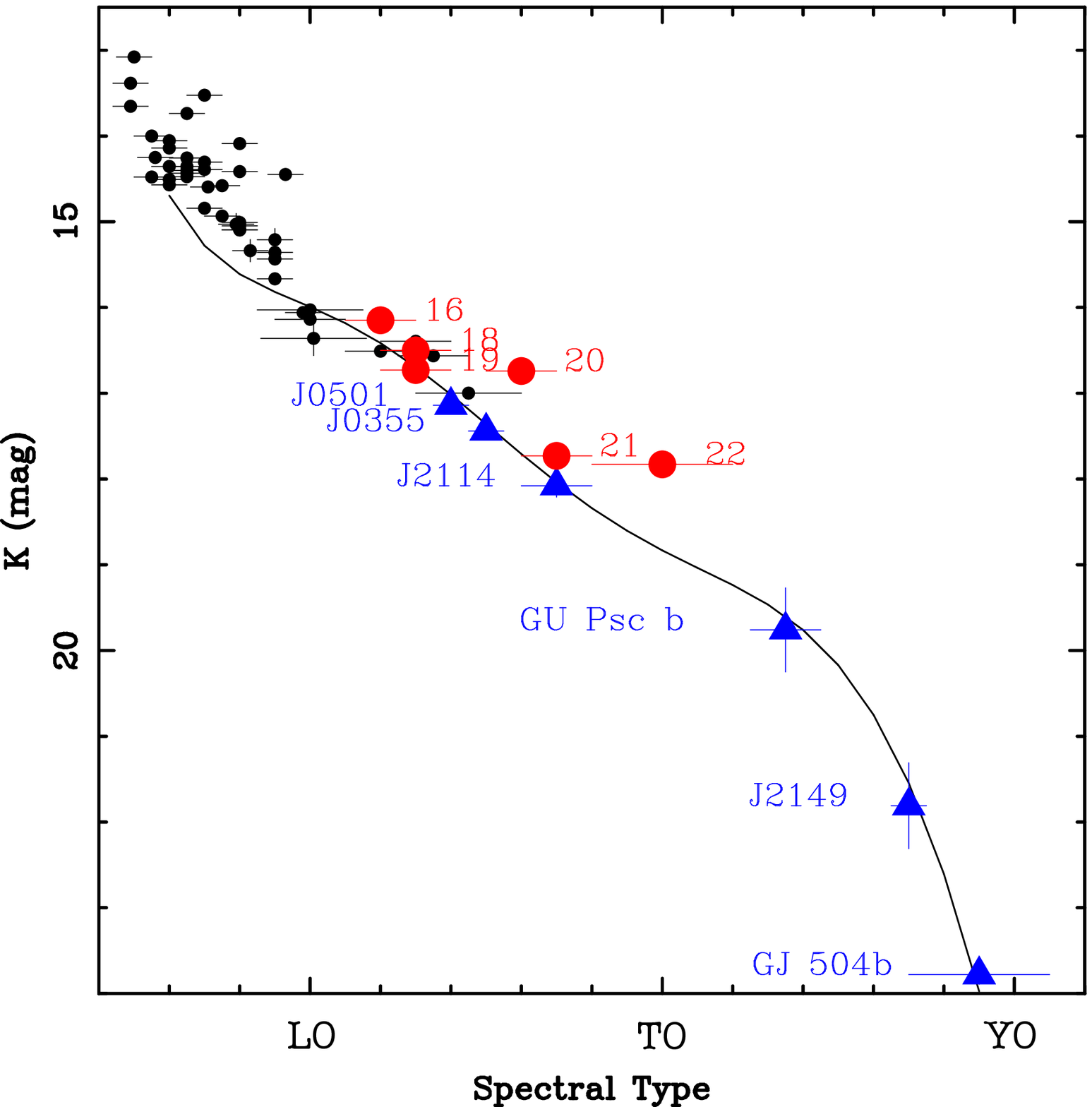}
   \caption{Left and  right panels depict the UKIRT $ZJK$ color-magnitude and spectral type-magnitude diagrams of confirmed Pleiades low-mass members (black and red dots). Calar~16,18--22, and~25 are labeled. Previously known members of \citet{bihain10} are plotted as black dots. The magenta symbols stand for \citet{osorio14a} candidates with $Z$ photometry extracted from UKIDSS (triangles and arrows) and \citet[square]{sarro14}. Masses for an age of 120 Myr are labeled in the left panel. The field L and T dwarfs with an age close to that of the Pleiades (\citealt{delorme12,liu13,kuzuhara13,faherty13,naud14,osorio14b}) are labeled and denoted by blue triangles. They are plotted as if they were located at 133.5 pc. Overlaid is the mean sequence of the field (solid line) produced by \citet{dupuy12} taken to the Pleiades distance.}
              \label{magspt}
    \end{figure}

The membership of our targets in the Pleiades was addressed in terms of the red optical-to-near-infrared colors, the late-type nature of the spectra, and the evidence of low-gravity atmospheres. Both $Z$-band photometry and near-infrared spectroscopy confirm that all seven objects are non-extragalactic sources and that they most likely have cool temperatures. The lefthand panel of Fig.~\ref{magspt} highlights the $K$ versus $Z-K$ color-magnitude diagram of Pleiades members. The new data are shown in red. Confirmed Pleiades objects from \citet{bihain10} and the bright cluster member candidates of \citet{osorio14a} are also included. Their $Z$ photometry was extracted from the UKIDSS database by cross-correlating the objects' coordinates against the archive within a searching radius of 1\arcsec. UKIDSS data extend down to $Z_{\rm{lim}}$\,$\approx$\,20.3 mag \citep{lodieu12}. The $Z$-band photometry of Calar~18 ($Z$ = 20.47\,$\pm$\,0.20 mag) was taken from the Sloan $z$ measurement of \citet{sarro14}, which was conveniently transformed into the Vega UKIDSS system by employing the color transformations given in \citet{hewett06}. Calar~21 nicely extrapolates the sequence of Pleiades low-mass members toward very faint magnitudes and red $Z-K$ colors. Its $Z-K$ index is compatible with the L7$\pm$1 classification (see next). We were able to impose the following lower limits on the colors of the faintest target, Calar~25: $Z-J \ge 2.52$ and $Z-K \ge 4.89$ mag. They indicate its extremely red optical-to-near-infrared slope typical of $\ge$mid-L ultra-cool dwarfs \citep{hewett06}, as opposed to very red high redshift galaxies, which appear bluer at these indices, particularly $Z-J$ (see \citealt{francis97,bihain09}). Calar~25 thus becomes our most promising least massive Pleiades proper-motion candidate that awaits spectroscopic follow-up. We predict a spectral classification probably within the late-L and T (methane) types. 

We determined the spectral types of the LIRIS and NIRSPEC data via visual comparison of the overall shape of the observations with spectra of known dwarfs taken with the same instrumentation or publicly available from previous spectroscopic works (Sect.~\ref{appendix}). This procedure yields a spectral classification mainly based on the spectral slope (or flux ratios) and the intensity of the water vapor absorption bands present throughout; the latter shows an almost linear behavior with spectral type from M6 through T6 \citep{mclean03}. All of our spectra display signatures indicative of cool dwarfs belonging to the L (and possibly early-T) domain \citep{kirk99,martin99}: Water-vapor absorption appears intense at 1.33 $\mu$m and in the $H$ and $K$ bands in all spectroscopic targets, which represents the most direct evidence of a cool nature. There is a markedly increasing slope from the blue to near-infrared wavelengths in the $ZJ$ spectra of Calar~16 and 18--20 consistent with early- to mid-L types, and carbon monoxide absorption at 2.30 $\mu$m appears to be detected in Calar~16, 20, and 21, despite the low resolution and poor S/N of the observed spectra. Both LIRIS and NIRSPEC spectra of Calar~21 show a flux decrement at around 2.3 $\mu$m consistent with the expected position of the CO feature. The high intensity of water vapor and the $H$/$K$ flux ratio indicate spectral types typical of mid- to late-L for Calar~20 and~21. The $K$-band spectrum of Calar~22 shows a decreasing flux redward of $\sim$2.15 $\mu$m, which we might assign to the onset of methane absorption defining the boundary between L and T types (\citealt{mclean03} noted, however, that CH$_4$ may appear in the $K$-band as early as L7). Additionally, the $H$ and $K$ steam bands seem stronger in Calar~22 than in Calar~21 and~20, thus suggesting a cooler classification for the former object. We tentatively classified Calar~22 as an L/T transition object. Our final spectral type assignments and their uncertainties are given in the last column of Table~\ref{obslog}. Spectral type error bars were obtained as the approximate cool and warm subtypes at which the observed spectroscopic slopes start to deviate slightly from the reference objects. In this respect, the classification assignments given in Table~\ref{obslog} should be understood most likely as spectral intervals rather than fixed values.

To find evidence of the low-gravity atmospheres of our targets (a signpost of young ages), we plot the LIRIS and NIRSPEC data in comparison with high- and low-gravity (field) dwarfs of related classification in the left- and righthand panels of Fig.~\ref{spectra}. Among the young (low-gravity) reference sources, we selected G\,196$-$3B \citep{rebolo98}, 2MASS\,J00452143$+$1634446 \citep{wilson03}, 2MASS\, J035523.27$+$113343.7 \citep{reid08}, and 2MASS\,J12073347$-$3932540b \citep{chauvin04}, which have ages constrained in the interval 10--500 Myr by \citet{chauvin04}, \citet{faherty13}, and \citet{osorio14b}. All show clear optical and near-infrared spectroscopic features ascribed to low gravity (\citealt[and references therein]{mcgovern04,cruz09}). At the resolution and wavelength coverage of the data, there are few signatures that indicate low-pressure atmospheres that we may use to assess cluster membership. Given the poor quality of the LIRIS spectra, the K\,{\sc i} atomic lines at around 1.18 and 1.25 $\mu$m are not sampled or detected well and cannot be safely employed as the ultimate youth criterion. The K\,{\sc i} absorption lines appear weaker in low-gravity than in high-gravity L dwarfs (e.g., \citealt{mcgovern04}), which makes K\,{\sc i} detection even harder for young objects. We imposed upper limits of 6 \AA~(Calar~16), 8 \AA~(Calar~19), 10 \AA~(Calar~18), and 16 \AA~(Calar~20) on the pseudo-equivalent widths\footnote{Equivalent width measured with respect to a pseudo-continuum.} (pEWs) of each of the 1.25-$\mu$m K\,{\sc i} doublet lines. For their measured spectral types, the K\,{\sc i} 1.25-$\mu$m pEW upper limits of Calar~16 and~19 may provide marginal confirmation of their low-gravity atmospheres, since we determined pEW = 6--8 \AA~for L2--L4.5 field dwarfs, in agreement with the measurements listed in Table~7 by \citet{mclean03}. 

The peaked shape of the $H$-band flux \citep{lucas01} is, however, a significantly broader signature also associated with low- to intermediate-gravity atmospheres. According to \citet{osorio14b}, it can be present in low-gravity L-type atmospheres up to ages of 120--500 Myr. On the one hand, it can be easily recognized at very low spectral resolution and poor S/N. On the other hand, \citet{bihain10} point out that the peaked-$H$ band shape is not always present among early-to-mid L- type Pleiades dwarfs. The high flux level of the $K$ band compared to the $H$ band, compatible with very red $H-K$ colors that are not usually found among field L and T dwarfs, may also hint at low gravities (or higher metallicities). As seen from Fig.~\ref{spectra}, Calar~20, 21, and~22 show peaked $H$-band spectra, and their LIRIS $K$- versus $H$-band fluxes are reproduced better by young L5--L7 dwarfs than by high-gravity field objects. This provides support to the low-gravity nature of Calar 20--22. Calar~16 displays an $H$-band spectrum similar to that of field early-L dwarfs. The $HK$ spectra of Calar 18--19 are not available, and we cannot prospect their low-pressure atmospheres at these wavelengths.

The very red $Z-K$ colors of Calar~21 and~25 (left panel of Fig.~\ref{magspt}), in addition to the red near-infrared colors \citep{osorio14a}, support cloud opacity increasing at moderately low gravity and low temperature, as discussed by \citet{saumon08} and \citet{marley12}. The righthand panel of Fig.~\ref{magspt} displays the $K$-band spectral sequence of Pleiades low-mass members. This sequence spans over 6.5 mag in $J$ and 5 mag in $K$ (mid-M trhough L types). Our Pleiades objects nicely extrapolates the observed cluster trend to quite faint magnitudes. To put the Pleiades population into context, the field sequence of \citet{dupuy12} shifted to the cluster distance is overlaid in all panels of Fig.~\ref{magspt}. As already discussed by \citet{bihain06}, the near-infrared magnitudes of the field and the Pleiades overlap at these cool spectral types probably because L-type objects (including brown dwarfs and low-mass stars) with ages above $\sim$100 Myr tend to have very similar sizes (0.08--0.12 $R_\odot$) independently of mass and age (see discussion in \citealt{luhman12rev}). As in \citet{osorio14a}, we also added field L and T sources unrelated to the Pleiades but with ages estimated at around 120 Myr in the righthand panel of Fig.~\ref{magspt}. If their age is finally confirmed to be that of the Pleiades, these objects would provide a nice extrapolation of the 120-Myr sequence from the late-Ls down to the late-T types. Interestingly, the thus-built 120-Myr sequence appears to describe the same arrangement as the field sequence. Therefore, this must be the path that brown dwarfs and giant planets follow as they evolve before falling into the extreme optical and near-infrared faintness of the Y dwarf domain. 

We conclude that both the $Z$-band photometry and the $JHK$ spectroscopic observations confirm the cool nature and provide evidence of the young age of our targets (particularly Calar~16 and 19--22), which, combined with their proper motions and photometric properties, make them the coolest and faintest, genuine Pleiades members known to date. Their masses are estimated in the range $\sim$0.010--0.035 M$_\odot$ for solar metallicity and 120 Myr (based on the evolutionary models by \citealt{chabrier00}). Calar 21, 22, and~25, whose likely masses are $\sim$0.015, $\sim$0.015, and $\sim$0.010 M$_\odot$, lie very close to the deuterium-burning mass borderline. Calar~25 may become the first free-floating planet ever discovered and corroborated in the Pleiades. Based on these results, the Pleiades substellar mass function presented in \citet{osorio14a} does not need any strong correction, and it indicates the rising mass spectrum toward lower masses. These Calar objects are benchmarks for interpreting young and old ultracool brown dwarfs and giant planets in isolation and as companions to more massive sources.

\begin{acknowledgements}
We thank J$.$ Patience for sending us the electronic version of the spectrum of 2MASS\,J120733.47$-$393254.0b. We also thank the referee, G$.$ Basri, for his useful comments on this manuscript. Based on observations made with the William Herschel Telescope (WHT) operated on the island of La Palma by the Isaac Newton Group in the Spanish Observatorio del Roque de los Muchachos of the Instituto de Astrof\'\i sica de Canarias. We  thank the observing staff at the W$.$ M$.$ Keck Observatory that helped us with the observations and acquisition of data. We thank the Hawaiian ancestry who allowed us to observe from their sacred mountain. Support for Keck data analysis comes from an award issued by the  Jet Propulsion Laboratory (JPL),  National Aeronautics and Space Administration (NASA). This work is based in part on data obtained as part of the UKIRT Infrared Deep Sky Survey. This research has made use of the SIMBAD database, operated at CDS, Strasbourg, France. This work was financed by the Spanish Ministry of Economy and Competitiveness through the projects AYA2011-30147-C03-03 and AYA2010-20535, and by Sonderforschungsbereich SFB 881 "The Milky Way System" (subprogram B6) of the German Research Foundation. M$.$ C$.$ G$.$ O$.$ acknowledges the support of a JAE-Doc CSIC fellowship co-funded with the European Social Fund under the program {\em Junta para la Ampliaci\'on de Estudios}. E$.$ L$.$ M$.$ acknowledges a visiting appointment at the Geosciences department of the University of Florida and a Severo Ochoa senior visitor grant at the Instituto de Astrof\'isica de Canarias. 

\end{acknowledgements}

\bibliographystyle{aa} 
\bibliography{1.bib}


\Online

\begin{appendix}
\section{Observations and data reduction \label{appendix}}
We collected low-resolution near-infrared spectra of six faint Pleiades proper motion candidates from \citet{osorio14a} using the Long-slit Intermediate Resolution Infrared Spectrograph, LIRIS \citep{manchado04}, at the Cassegrain focus on the 4.2-m William Herschel Telescope (WHT) and the echelle and grating Near-Infrared Spectrometer, NIRSPEC \citep{mclean98}, at the Nasmyth focus on the 10-m Keck\,II telescope. LIRIS and NIRSPEC have 1024$\times$1024 HAWAII and ALADDIN-3 InSb detectors for the 0.8--2.5 and 0.95--5.5 $\mu$m ranges, respectively. The pixel scales are 0\farcs25 (LIRIS) and 0\farcs19 (NIRSPEC). Table~\ref{obslog} provides the journal of the spectroscopic observations including abridged targets names, observing dates in Universal Time (UT), integration times, and air masses. We selected the LIRIS $ZJ$ (0.9--1.35 $\mu$m) and $HK$ (1.45--2.4 $\mu$m) grisms and the NIRSPEC $K$-band (1.95--2.31 $\mu$m) low-resolution mode, which combined with slit widths of 1\arcsec~(LIRIS) and 0\farcs76 (NIRSPEC), that is four pixels along the dispersion directions of the detectors, provide final spectral dispersions and resolutions of 6.1~\AA\,pix$^{-1}$ and 25~\AA~($ZJ$), 9.8~\AA\,pix$^{-1}$ and  39 \AA~($HK$), and of 4.3~\AA\,pix$^{-1}$ and 17 \AA~($K$). The length of the slits were 4\arcmin~(LIRIS) and 42\arcsec~(NIRSPEC), which is long enough to simultaneously observe our targets and bright reference sources in the field of view. To align targets and their respective bright reference stars along the slits (all at less than 1\arcmin-distance from the Pleiades sources), we conveniently rotated the instruments. The weather during the WHT campaigns were photometric, and cirrus were present during the Keck observations. Raw seeing oscillated between 0\farcs9 and 1\farcs2. 

Spectra were acquired at two different nodding positions on the detectors for a proper sky subtraction. Raw data were reduced using standard procedures for the near-infrared wavelengths, including sky subtraction and flat-fielding correction, and packages within the IRAF\footnote{The Image Reduction and Analysis Facility (IRAF) is distributed by the National Optical Astronomy Observatories, which are operated by the Association of Universities for Research in Astronomy, Inc., under cooperative agreement with the National Science Foundation.} environment. Individual flat-fielded, sky-subtracted frames were properly registered using the bright reference stars and stacked to produce deep data. Spectra of the Pleiades targets were optimally extracted using the IRAF APALL task and were calibrated in wavelength to precisions of $\pm$0.3--0.5 \AA~using arc lines of Ne$+$Ar. The spectra of hot B and early-A stars observed immediately after our targets were used for division into the corresponding science spectra. We previously removed the hydrogen lines intrinsic to these hot stars. Finally, we multiplied the science spectra by the blackbodies of temperatures typical of B/A stars. Figure~\ref{spectra} illustrates the resulting LIRIS and NIRSPEC spectra of Pleiades candidates. Given the faint magnitudes of our spectroscopic targets ($J$\,=\,17.5--20.3 and $K$\,=\,16.1--17.8 mag), some spectra have a poor signal-to-noise ratio (S/N = 2.5--8). To improve the data quality, we depict binned spectra as indicated in the figure caption. For a proper scaling of the $ZJ$ and $HK$ separate spectra of Calar~16 and~20, we employed the objects' $J$ and $H$ photometry published in \citet{osorio14a}.

We also acquired Sloan $z$-band images of Calar~21 and~25 using the Auxiliary-port Camera (ACAM) at the folded-Cassegrain focus on the WHT. Table~\ref{obslog} provides the log of the imaging observations. ACAM has a 2048$\times$4096 EEV CCD detector with a pixel pitch of 0\farcs25 on the sky. The ACAM observations of 2014 February were conducted under photometric conditions and a seeing of 1\farcs2 ($z$), while there were thin cirrus and variable seeing during the 2014 January campaign. The $z$-band data were taken following a nine-point dither pattern for a proper sky subtraction. Groups of three to nine images were combined to produce the mean sky contribution every 3 to 18 min, which were later removed from the raw data. Images were corrected for flat-field, properly aligned, and those with the best seeing were stacked to deliver one final deep image per target. Aperture and point-spread-function photometry were obtained within the IRAF PHOT package. Photometric calibration was performed using the $Z$-band data of bright sources in the field of view of our targets provided by the UKIRT Infrared Deep Sky Survey (UKIDSS, \citealt{lawrence07}), which fully overlaps with the Pleiades. We list the measured $Z$-band magnitudes in Table~\ref{obslog}; the Sloan magnitudes were converted into the UKIDSS system using the AB to Vega correction and the color terms appropriate for mid- to late-L types given by \citet{hewett06}. Calar~25 remains undetected after 1.4-h on source integration time.  In Table~\ref{obslog}, we list a 4-$\sigma$ $Z$-band upper limit of 23.35 mag ($\sim$3 mag deeper than the UKIDSS survey), where $\sigma$ stands for the sky noise at the position expected for the source.

Together with the Pleiades candidates, we also observed known spectroscopic and photometric L-type reference dwarfs by employing the same LIRIS and ACAM instrumental configurations as for the targets. Table~\ref{obslog} contains the journal of these observations. All raw data were reduced in the same manner as the Pleiades candidates. The sources 2MASS\,J22244381$-$0158521 and 2MASS\,J09083803$+$5032088 are field L4.5 and L5 dwarfs \citep{kirk00,cruz03}. The latter has Sloan $z$ and UKIDSS $Z$ photometry, which allowed us to confirm the color-term correction between the two photometric systems valid for L5 dwarfs. The objects 2MASS\,J00452143$+$1634446 (L2) and 2MASS\,J03552327$+$1133437 (L5) are two young field dwarfs with lithium detection at 670.8 nm and another spectroscopic features indicative of low-gravity atmospheres \citep{cruz09,faherty13,osorio14b}, whose ages are estimated at 10--100 and 50--500 Myr, respectively \citep{osorio14b}. The LIRIS spectra of the reference L dwarfs are plotted along with the Pleiades targets in Fig.~\ref{spectra}. Additional low-resolution spectra of DENIS\,J122815.2$-$154733 (L6), SDSS\,J042348.57$-$041403.5 (T0), and the young G\,196$-$3B (L3) and 2MASS\,J12073347$-$3932540b (L7) taken from the literature \citep{leggett01,patience10,osorio10} are also included in Fig.~\ref{spectra}.

\end{appendix}

\end{document}